# Determination of the spin Hall angle, spin mixing conductance and spin diffusion length in Ir/CoFeB for spin-orbitronic devices

T. Fache[1], J.C. Rojas-Sanchez[1*], L. Badie[1], S. Mangin[1] and S. Petit-Watelot[1†]

[1] *Université de Lorraine, CNRS, Institut Jean Lamour, F-54000 Nancy, France*

*Corresponding authors:*
*\* juan-carlos.rojas-sanchez@univ-lorraine.fr*
*† sebastien.petit-watelot@univ-lorraine.fr*

Iridium is a very promising material for spintronic applications due to its interesting magnetic properties such as large RKKY exchange coupling as well as its large spin-orbit coupling value. Ir is for instance used as a spacer layer for perpendicular synthetic antiferromagnetic or ferrimagnet systems. However, only a few studies of the spintronic parameters of this material have been reported. In this paper, we present inverse spin Hall effect - spin pumping ferromagnetic resonance measurements on CoFeB/Ir based bilayers to estimate the values of the effective spin Hall angle, the spin diffusion length within iridium, and the spin mixing conductance in the CoFeB/Ir bilayer. In order to have reliable results, we performed the same experiments on CoFeB/Pt bilayers, which behavior is well known due to numerous reported studies. Our experimental results show that the spin diffusion length within iridium is 1.3 nm for resistivity of 250 nΩ.m, the spin mixing conductance $g_{eff}^{\uparrow\downarrow}$ of the CoFeB/Ir interface is 30 nm$^{-2}$, and the spin Hall angle of iridium has the same sign than the one of platinum and is evaluated at 26% of the one of platinum. The value of the spin Hall angle found is 7.7% for Pt and 2% for Ir. These relevant parameters shall be useful to consider Ir in new concepts and devices combining spin-orbit torque and spin-transfer torque.

## INTRODUCTION

Iridium is a very promising material for spintronic applications. Its properties include large spin orbit coupling [1], large Ruderman-Kittel-Kasuya-Yosida (RKKY) exchange coupling [2], and strong interface contribution to perpendicular magnetic anisotropy (PMA) [3–6] and to interfacial Dzyaloshinskii-Moriya interaction [7]. Ir has been shown to be a key element as a spacer layer to create model perpendicular synthetic ferrimagnets or synthetic antiferromagnets [8,9]. However, to our knowledge, the spin Hall effect, which has proven to be an efficient physical effect to manipulate magnetisation [10–16], was scarcely studied in Ir [7,17,18]. Given the scattering among the spin transport measurements, and spin-Hall effect (SHE) characterisations, we have decided to perform a comparative study. Indeed, in this paper, we display the results of spin pumping voltage induced ferromagnetic resonance (SP–FMR) experiments obtained on Co$_{40}$Fe$_{40}$B$_{20}$/Ir and Co$_{40}$Fe$_{40}$B$_{20}$/Pt based bilayers, as it has been proven that SP-FMR is one of the most effective experiments in order to probe the inverse spin-Hall effects (ISHE) [19–21]. We characterised the SHE for iridium layers in comparison to Pt, which spin Hall behaviour is already well-known [22–35]. Therefore, we propose here a comparative approach to determine the spin-to-charge current conversion efficiency. We present our experimental determination of the spin diffusion length ($l_{sf}$), the effective spin mixing conductance ($g_{eff}^{\uparrow\downarrow}$) and the effective spin Hall angle ($\theta_{SHE}$) for iridium based materials.

## SAMPLES GROWTH AND SP-FMR MEASUREMENTS

The samples used in our experiments are Si-SiO$_2$(300 nm)//CoFeB(5 nm)/Ir and Si-SiO$_2$(300 nm)//CoFeB(5 nm)/Pt bilayers deposited on thermally oxidized silicon substrates. The double slash, //, stands for the position of the substrate. We chose to grow samples with various iridium or platinum capping thickness (1, 2, 4, 6 and 15 nm) but keeping the same 5 nm Co$_{0.4}$Fe$_{0.4}$B$_{0.2}$ ferromagnetic layer thickness. All the layers were deposited by magnetron sputtering with a base vacuum pressure of 8×10$^{-9}$ mbar, and an argon deposition pressure of 5×10$^{-3}$ mbar, from pure elemental targets for platinum and iridium, and from an alloy Co$_{0.4}$Fe$_{0.4}$B$_{0.2}$ target. The samples grown were then patterned into rectangular slabs with electrodes at the end to measure the spin pumping voltage. The devices were structured using standard optical lithography techniques. An antenna was patterned on top of the sample, which was insulated by a SiO$_2$ cap of 75 nm. In this antenna we injected a GHz RF current which generates the radio-frequency field $h_{rf}$ on the sample. A static magnetic field ranging from 0 to 0.5 T can also be applied in plane. Depending on the value of the applied field, $h_{rf}$ could excite the magnetisation at resonance. Through the spin-pumping effect, the precession of the magnetisation yield to the creation of a pure transverse spin current $j_s$ that is injected from the CoFeB layer into the Ir or Pt layer. $j_s$ is generally expressed as follows [19]:

$$j_S = \frac{\hbar Re(g^{\uparrow\downarrow})}{4\pi} \int_0^{\frac{2\pi}{\omega}} \left( \boldsymbol{m} \times \frac{d\boldsymbol{m}}{dt} \right) dt \quad (1)$$

Where $g^{\uparrow\downarrow}$ is the spin mixing conductance, $\boldsymbol{m}$ is the reduced magnetization and ω is the RF field pulsation. This spin current injected in the Heavy Metal (HM) layer is then converted into a charge current due to the ISHE which is detected by electrical means, by measuring the voltage in an open circuit, as represented in Fig. 1. Such a voltage due to spin pumping is symmetrical around the resonance field as shown in Fig 2.b. From the peaks at resonance, several materials properties can be deduced. For instance, from this ISHE spin pumping voltage, the resonance magnetic field, $H_{res}$, as a function of the frequency $f$ of the RF excitation applied to the sample gave us information about the ferromagnetic layer excited.

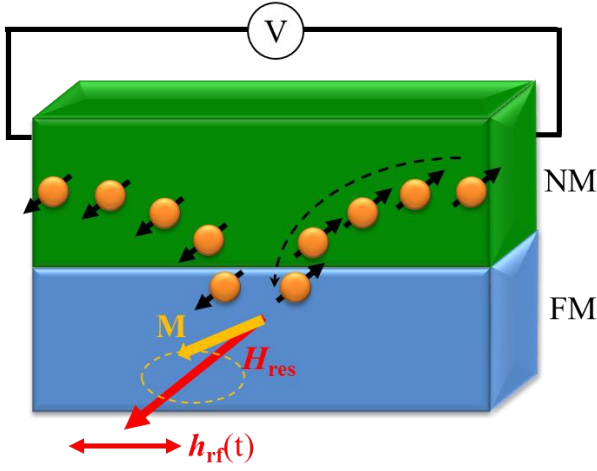

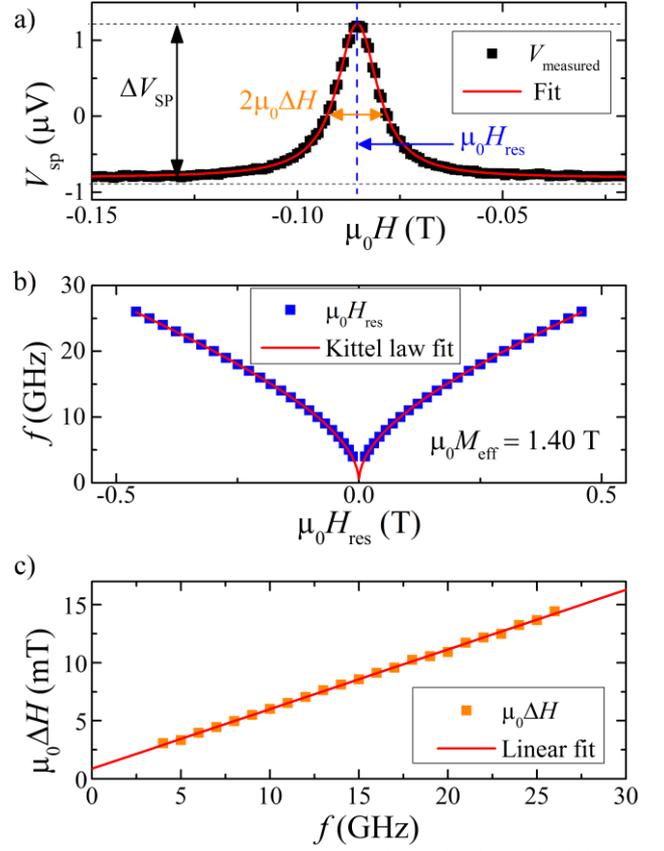

*Figure 1: Schematic of the bilayer sample composed of a ferromagnetic layer of CoFeB (FM) and a non-magnetic material NM. The static field $H_{res}$ is applied perpendicularly to the wire, whereas the RF field $h_{rf}$ is aligned with the wire's length. The magnetisation M precesses around $H_{res}$, generating a spin current polarised along $H_{res}$, and which is injected in the NM layer. This injected spin current in the NM layer is converted into a charge current trough the ISHE. In an open circuit we can detect the voltage V as depicted. The charge current production is nothing else that the voltage amplitude normalized by the total resistance of the sample.*

The relationship between these parameters is described by the Kittel law [36], as presented on Fig. 1c:

$$f = \frac{\gamma \mu_0}{2\pi}\sqrt{H_{res}(H_{res}+M_{eff})} \quad (2)$$

where $\gamma$ is the gyromagnetic ratio, $\mu_0$ is the vacuum permeability, and $M_{eff}$ is the effective saturation magnetisation of the FM layer. $M_{eff}$ that is extracted does not change significantly with the thickness of the non-magnetic material (here Pt or Ir) that caps the ferromagnetic layer. For all the samples grown, the value of $\mu_0 M_{eff}$ is evaluated at 1.39 T.

*Figure 2: Measurements on //CoFeB(5 nm)/Ir(4 nm) (a) Spin pumping voltage at a frequency of 10 GHz ; (b) Linewidth ($\mu_0 \Delta H$) as a function of the frequency; (c) frequency ($f_r$) vs. the applied magnetic field ($\mu_0 H$). On all these graphs, symbols represent experimental measurements whereas solid red lines correspond to fits. In (c) the Kittel law, given by equation 2, yields $\mu_0 M_{eff}$ =1.40 T.*

On the other hand, from the measurement of the spin pumping voltage as a function of the applied field (Fig. 2.a) one can determine the linewidth $\Delta H$ of the voltage peaks to estimate the magnetic damping $\alpha$ of the materials. The evolution of the linewidth as a function of the frequency (Fig. 2.c) of the excitation shows a linear dependence, which slope is found to be proportional to the effective damping, following the relationship [37]:

$$\mu_0 \Delta H = \mu_0 \Delta H_0 + \frac{2\pi\alpha}{\gamma}f \quad (3)$$

where $\Delta H$ is the linewidth of the ferromagnetic resonance peak and $\Delta H_0$ is the inhomogeneous broadening. An example of such damping determination is shown in Fig. 2c. We have performed such a damping determination on both series of samples. Figure 3 shows the evolution of the effective damping $\alpha$ as a function of the thickness of nonmagnetic material:

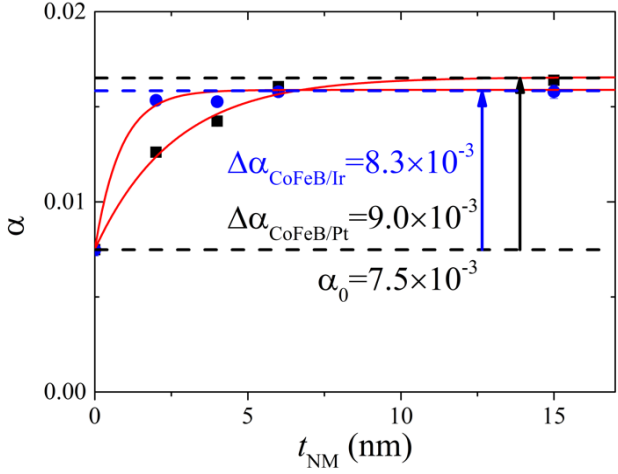

*Figure 3: Evolution of the deduced magnetic damping α from SP-FMR as a function of the thickness of the nonmagnetic material (Ir or Pt) on //CoFeB(5nm)/ Ir or Pt ($t_{NM}$ nm). In the case where $t_{NM} = 0$, the magnetic damping value is obtained from FMR measurements. Red curves stands for an exponential decay, $\alpha = \Delta\alpha(1 - e^{-t_{NM}/l_d})$. It results that $l_d$ =2.6±0.3 (0.9±0.2) nm for CoFeB/Pt (CoFeB/Ir). Note that there is not a factor 2 in the exponential argument.*

To measure the value of the intrinsic damping we used a reference sample of CoFeB (5 nm) capped with Al (3nm). The intrinsic damping value $\alpha_0 = 7.5 \times 10^{-3}$ for $t_{NM} = 0$ was obtained by Vector network analyzer (VNA)-FMR, since SP-FMR could not be an efficient detection method in that special case. Moreover, for $t_{NM}>0$ we have checked that the damping values obtained using both methods were consistent. The value of the intrinsic damping was found to be slightly higher than the values given in part of the literature (4–5 ×10$^{-3}$) [38–41]. This could be explained by the growth conditions. Indeed, the studied carried out by Xu *et al.* [42] shows that the magnetic damping of sputtered CoFeB is very sensitive and decreases when the argon pressure increases. They reported a damping value of 13×10$^{-3}$ when Ar pressure was 3 mTorr. The annealing also affects the damping [38,41]. Another possibility might be to attribute this slight difference to the aluminium oxide layer that caps the CoFeB magnetic layer probed by FMR in order to obtain the intrinsic damping.

As we can see in Fig. 3, the magnetic damping increases strongly for both capping layers, Pt as well as Ir. This phenomenon is a well-known feature of damping enhancement due to spin pumping effect [19,26,29], and can be characterised by the following relation:

$$\Delta\alpha = \alpha - \alpha_0 = \frac{g\mu_B g_{eff}^{\uparrow\downarrow}}{4\pi M_S t_{FM}} \quad (4)$$

where g is the Landé factor (2.11 for CoFeB), $g_{eff}^{\uparrow\downarrow}$ is the real part of the effective spin mixing conductance, $\mu_B$ is the Bohr magneton and $t_{FM}$ is the ferromagnetic (CoFeB in our case ) layer thickness . Replacing our experimental values in eq. (4), we estimated the value of $g_{eff}^{\uparrow\downarrow}$ for the Ir/CoFeB interface to be around 30 nm$^{-2}$, and 32nm$^{-2}$ for the Pt/CoFeB interface. These values are in the typical order of magnitude of effective spin mixing conductances obtained for ferromagnetic/Pt systems (for Py/Pt: 21 to 30 nm$^{-2}$ [23,29,31,33]; for Co/Pt: 80nm$^{-2}$ [26]) and epitaxial Fe/Pt: 26 nm$^{-2}$ [43]. Especially, for the CoFeB/Pt interface, some reported values are 40 nm$^{-2}$ [44], 54 nm$^{-2}$ and 47 nm$^{-2}$ for the opposite stacking order, Pt/CoFeB, in [45] and 50.7 nm$^{-2}$ in [46]). In the case of Ir it has been reported for NiFe/Ir interfaces so far, 13 nm$^{-2}$ in [17] and 25.2 nm$^{-2}$ in [18]. Those values are effective value since they include interface contributions such a spin memory loss [18,26]. Let us point out here that we do not consider the imaginary part of the spin mixing conductance in our work, since the Kittel fittings performed do not show a value of the gyromagnetic ratio differing from the one of electrons, as predicted for metallic systems [19,47].

**SPIN DIFFUSION LENGTH AND SPIN HALL ANGLE DETERMINATION** .

Finally, the spin pumping voltage measured normalized by the resistance of the FM/NM slab gives the charge current produced by ISHE. In this geometry the charge current measured can be expressed as follows [26,27,29]:

$$I_C = \theta_{SHE} l_{sf} w J_S^{eff} \tanh\left(\frac{t_{NM}}{2l_{sf}}\right) \quad (5)$$

where $w = 10$ μm is the width of the device. Here, $J_S^{eff}$ is the effective spin current density injected in the NM layer and it follows the relationship [26,27]:

$$J_S^{eff} = \frac{e g_{eff}^{\uparrow\downarrow} \gamma^2 h_{RF}^2}{4\pi\alpha^2} A(\omega) \quad (6)$$

where $A(\omega) = \frac{\gamma\mu_0 M_{eff} + \sqrt{(\gamma\mu_0 M_{eff})^2 + 4\omega^2}}{(\gamma\mu_0 M_{eff})^2 + 4\omega^2}$ represents the influence of the magnetic dynamics on the injected spin currents, as it was shown by Ando *et al* [27]. Figure 4 shows the raw data of spin pumping voltage ($V_{sp}$) as a function of the applied magnetic field (μ$_0$H) for an excitation frequency of 15 GHz. We can observe that the voltage is a purely symmetric lorenztian around resonance field and it changes its sign upon changing the sign of the applied field. All these are features of an ISHE spin pumping voltage. Furthemore, that is also verified in the case of CoFeB/Pt bilayer.

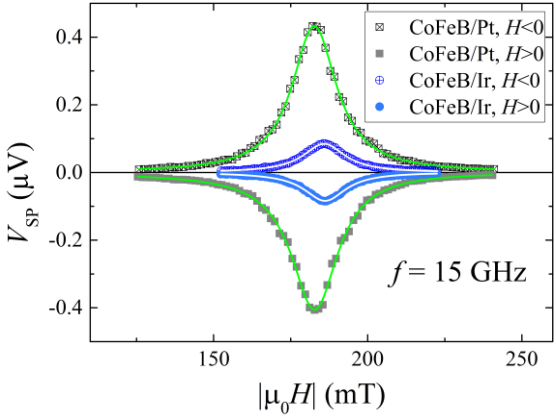

Figure 4: Spin pumping voltage $V_{sp}$ for Si/SiO$_2$/ CoFeB (5nm)/Ir (6nm) and Si/SiO$_2$/CoFeB(5nm)/Pt(6nm) as a function of the applied field absolute value (|H|) for an excitation frequency of 15 GHz. Results are shown for positive and negative static applied fields. Symbols represent experimental measurements whereas the full lines correspond to a fit of the data by a Lorentzian function. A constant offset was subtracted. The raw voltage is purely symmetrical, getting rid of the spurious signals.

Figure 5 shows the evolution of the charge current produced as a function of the iridium thickness in CoFeB/Ir bilayers for various frequencies. Using these results, the spin diffusion length for Iridium, $l_{sf}^{Ir}$, can be deduced from equation (5). Thus, $l_{sf}^{Ir}$ obtained for each frequency is displayed in Figure 6. Our results show consistent values of $l_{sf}^{Ir} = 1.3 \pm 0.1$ nm, and $l_{sf}^{Pt} = 2.4 \pm 0.3$ nm (red dashed lines).

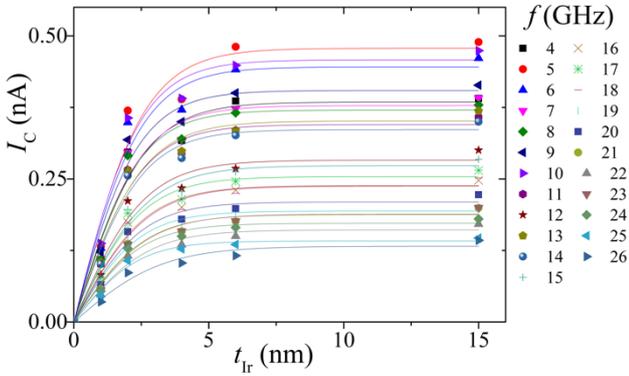

Figure 5: Produced charge current ($I_c$) as a function of the iridium thickness ($t_{Ir}$) for frequencies ranging from 4 to 26 GHz. The symbols represent experimental values whereas the solid lines show the fitting obtained thanks to eq. (5)

The value of spin diffusion length obtained for platinum is in agreement with the values found in the literature: the experimental values reported using SP-FMR set-up range from 0.5 nm [32], to 10 nm [23], with numerous values in between [10,48]. The obtained value $l_{sf}^{Ir} = 1.3$ nm is twice larger than the one presented in earlier studies [17,18] with similar FMR-based methods, and close to the one reported by spin-orbit torque technique, ~1 nm [7]. That difference might be due to different Ir resistivity. However, we would like to note that $l_{sf}$ values reported only by spin pumping FMR measurements (not spin pumping voltage measurements) consider an exponential decay of damping with and argument ($2l_{sf}/t_{NM}$). However, the $t_{NM}$ damping evolution is not reliable to estimate $l_{sf}$ as it was pointed out in ref. [26]. We can observe that discrepancy with results in Fig. 3 where $l_d$ is close to $l_{sf}$ estimated by charge current dependence in Fig. 5 and 6 but we have used a different exponential argument ($l_d/t_{NM}$). This is likely to explain the difference with the two previous studies [17,18].

Further, $l_{sf}^{Ir} = 1.3$ nm can be compared to the usual range of thicknesses where iridium is used, especially in the case of synthetic ferrimagnets where the iridium spacer is used to maximize the RKKY coupling, around 0.5 nm or 1.5 nm (1$^{st}$ and 2$^{nd}$ peaks) [2,8,9]. From the experimental values of spin diffusion length and resistivity of the HM layer, we can compute its spin resistance $r_s = \rho \times l_{sf}$. The resistivities measured for Pt and Ir are the following: $\rho_{Pt} = 245$ nΩ.m and $\rho_{Ir} = 250$ nΩ.m. We thus have the spin resistance $r_{s,Pt} = 0.59$ fΩ.m$^2$ and $r_{s,Ir} = 0.32$ fΩ.m$^2$. The value of $r_{s,Pt}$ is very close to the experimental result published in ref. [26] as well as close to the theoretical value reported by Liu et al. [49]. We can also use the remark from reference [26], stating that in the case of Pt, given the results reported in the literature, the product of the effective spin Hall angle and the spin diffusion length, $\theta_{SH} \times l_{sf}$, is a quantity that is nearly independent on the technique or the setup used, and its effective value is estimated to be close to 0.19 nm. It is therefore possible to obtain the effective spin Hall angle of platinum, leading to a value of $\theta_{SH}^{Pt} \approx 7.6\%$.

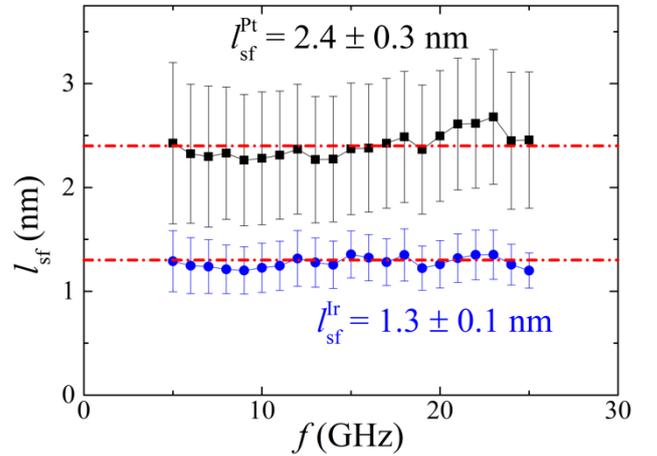

Figure 6: Spin diffusion length $l_{sf}$ as a function of the frequency in iridium and platinum deduced from the fit shown in figure 5.

To determine $\theta_{SH}^{Ir}$ accurately and independently, the value of the effective spin current is needed. In order to do so, it is mandatory to estimate the strength of the radio frequency excitation field, $h_{rf}$, and its frequency dependence, as well as the $g_{eff}^{\uparrow\downarrow}$ factor for the CoFeB/Ir interface. The latter was previously estimated to 30 nm². After accurate measurements of the transmission line and of the scattering matrices of the devices corresponding to our samples, we have concluded that the frequency dependence of $h_{RF}$ with respect to the frequency of the signal is the same for the iridium and the platinum based samples. This was expected, since the values of the conductivities are found to be very close for both materials.

Now, we defined the quantity:

$$\mathfrak{I}_{SP} = \frac{I_C \cdot \alpha^2}{l_{sf} \cdot g_{eff}^{\uparrow\downarrow} \cdot A(\omega)} \quad (7)$$

We can then plot the ratio $\frac{\mathfrak{I}_{SP}^{Ir}}{\mathfrak{I}_{SP}^{Pt}}$ of this $\mathfrak{I}$ parameter given in eq. (7) for different frequencies experimentally measured as displayed in Fig. 7. This $\frac{\mathfrak{I}_{SP}^{Ir}}{\mathfrak{I}_{SP}^{Pt}}$ value is most likely to give the right estimate of the actual spin Hall angles ratios, since it can be interpreted as:

$$\frac{\mathfrak{I}_{SP}^{Ir}}{\mathfrak{I}_{SP}^{Pt}} = \frac{\theta_{SH}^{Ir}}{\theta_{SH}^{Pt}} \frac{\tanh(t_{Ir}/(2l_{sf}^{Ir}))}{\tanh(t_{Pt}/(2l_{sf}^{Pt}))} \quad (8)$$

The limit obtained for $t_{NM} \gg l_{sf}^{NM}$ is the ratio of spin Hall angles. Indeed, we can represent this ratio as a function of the nonmagnetic materials thickness as shown in Fig. 8. We can observe a very large discrepancy between the value of the ratio given in eq. (8) and the one obtained experimentally for $t_{NM} = 2$nm.

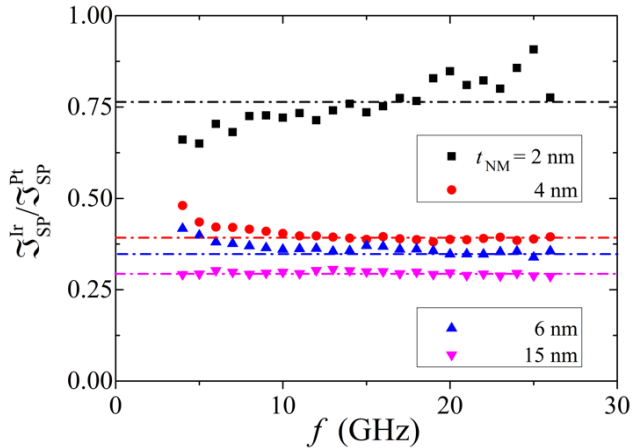

Figure 7: ratio $\mathfrak{I}$ of the spin pumping currents based on eq.( 7) as a function of the frequency. The dashed lines are guidelines towards the value obtained at high frequency.

Numerous elements can explain the difference between the model given in eq. (8) and the experimental results at low thickness. First, we can question the validity of the assumptions used in our study. We have considered that the resistivity, the spin diffusion length, and the spin Hall angle of the materials were independent of the nonmagnetic material thickness. However, this approximation does not hold for very low thicknesses, which is where the model and the experimental results do not match. Furthermore, at very low thicknesses, the roughness and the quality of the interface plays a larger role than for thick layers. The errors on the thicknesses and on the ratios are expected to be larger than for thicker samples.

Nevertheless, a good agreement for nonmagnetic materials thickness superior to 2nm is obtained, and we can estimate the ratios of effective SHE efficiencies to be $\frac{\theta_{SH}^{Ir}}{\theta_{SH}^{Pt}} = 0.26$. This approach lets us evaluate the values of $l_{sf}$ and $\theta_{SH}$ with precision.

Besides, using our determination of the spin Hall angle of platinum at 7.6%, we can estimate the spin Hall angle of iridium to be around 2%. Literature provides a large range of values for Pt that span to more than an order of magnitude, ranging from 0.33 to 0.0067 [22,23,32–35,47,48,24–31]. Our result for Ir is in good agreement with what was found in ref. [17], with a 2% value, and twice the one reported by spin-orbit torque in ref. [7]. The method that we present here enables to make a comparison by getting rid of many artefacts that seem to be the cause of a broadening of the results obtained in the literature.

We can use the works in ref. [50] to evaluate the efficiency of these two materials for spin-to-charge and charge-to-spin conversion applications. For the generation of a charge current, the figure of merit proposed is the product $\lambda_{ISHE}^* = \theta_{SH} \times l_{sf}$. We find a value of 0.186 nm for Pt, and 0.026 nm for Ir, suggesting that iridium is a poor candidate for further spin pumping applications. However, if we consider the figure of merit to assess the spin current generation, which is mandatory for spin orbit torque (SOT), given by the formula $q^* = 0.38 \times \frac{\theta_{SH}}{l_{sf}}$, we find a value of $12\times10^{-3}$ nm$^{-1}$ for Pt and $\sim 6\times10^{-3}$ nm$^{-1}$ for Ir. Therefore, it appears that even though Pt is the best material amongst those studied in both cases, Ir is good candidate, with half the ability of Pt to generate efficiently a spin current.

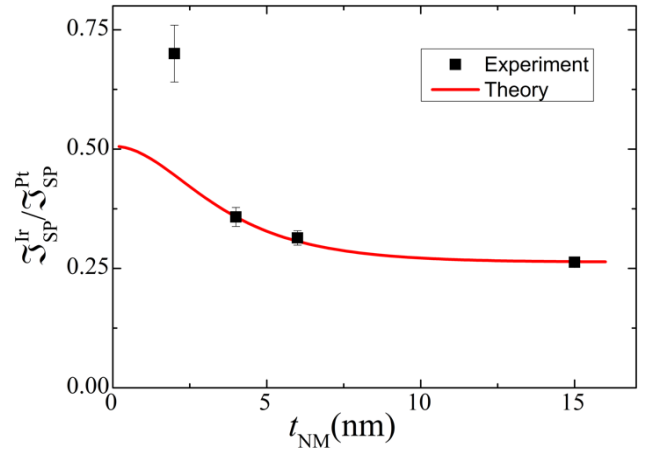

Figure 8: Evolution of the corrected spin pumping currents ratio as a function of the non-magnetic materials thickness (Black squares) and the expected dependence (red line) according to eq. (8).

## CONCLUSION

In this paper, we have described an approach that enables the measurement of the spin Hall angle of a material with respect to another one. We report reliable values of spin diffusion lengths of $1.3 \pm 0.1$ nm for iridium, $2.4 \pm 0.3$ nm for platinum from the NM thickness dependence of the charge current (and not from damping evolution). The spin mixing conductances for both interfaces CoFeB/Ir and CoFeB/Pt have been estimated around 30 and 32 nm$^{-2}$, respectively. The spin Hall angle of Ir has the same sign as the one of Pt and represents 26% of its value. We could obtain a $\theta_{SH}$ value of 7.6% for Pt, from which we could deduce a $\theta_{SH}$ value of 2% for Ir. Even though this procedure does not give by itself the value of the spin Hall characteristics of a material, it gives information about materials in same conditions, and enables a comparison between various materials. This can be an opportunity to unify the results concerning spin diffusion lengths and spin Hall angles, given the large dispersion in the results reported in the past decade. The spintronic parameters we are reporting for Ir will appeal for more applications exploiting this material in new spin-orbitronic devices such as combined spin-orbit torque and spin transfer torque effects in magnetic tunnel junctions [51] . This is by combining two major effects in spintronics, RKKY and SHE.

## AKNOWLEDGEMENTS

T. F. thanks the ANRT and the company Vinci Technologies for funding his PhD, under the CIFRE convention No. 2016/1458. All authors acknowledge support from Agence Nationale de la Recherche (France) under contract N° ANR-18-CE24-0008 (MISSION) and ANR-19-CE24-0016-01 (TOPTRONIC), from the French PIA project "Lorraine Université d'Excellence", reference ANR-15IDEX-04-LUE, from Region Grand Est, Metropole du Grand Nancy, Institut Carnot ICEEL, from the "FEDER-FSE Lorraine et Massif Vosges 2014-2020", a European Union Program, and from. Devices in the present study were patterned at MiNaLor clean-room platform which is partially supported by FEDER and Grand Est Region through the RaNGE project.

## REFERENCES

[1] J. P. Clancy, N. Chen, C. Y. Kim, W. F. Chen, K. W. Plumb, B. C. Jeon, T. W. Noh, and Y. J. Kim, "Spin-orbit coupling in iridium-based 5d compounds probed by x-ray absorption spectroscopy" *Phys. Rev. B - Condens. Matter Mater. Phys.* **86**, 195131 (2012).

[2] S. S. P. Parkin, "Systematic variation of the strength and oscillation period of indirect magnetic exchange coupling through the 3 d , 4 d , and 5 d transition metals" *Phys. Rev. Lett.* **67**, 3598 (1991).

[3] K. Nakamura, T. Nomura, A.-M. Pradipto, K. Nawa, T. Akiyama, and T. Ito, "Effect of heavy-metal insertions at Fe/MgO interfaces on electric-field-induced modification of magnetocrystalline anisotropy" *J. Magn. Magn. Mater.* **429**, 214 (2017).

[4] P. Taivansaikhan, D. Odkhuu, S. H. Rhim, and S. C. Hong, "Gigantic perpendicular magnetic anisotropy of heavy transition metal cappings on Fe/MgO(0 0 1)" *J. Magn. Magn. Mater.* **442**, 183 (2017).

[5] T. Nozaki, A. Kozioł-Rachwał, M. Tsujikawa, Y. Shiota, X. Xu, T. Ohkubo, T. Tsukahara, S. Miwa, M. Suzuki, S. Tamaru, H. Kubota, A. Fukushima, K. Hono, M. Shirai, Y. Suzuki, and S. Yuasa, "Highly efficient voltage control of spin and enhanced interfacial perpendicular magnetic anisotropy in iridium-doped Fe/MgO magnetic tunnel junctions" *NPG Asia Mater.* **9**, e451 (2017).

[6] Y.-C. Lau, Z. Chi, T. Taniguchi, M. Kawaguchi, G. Shibata, N. Kawamura, M. Suzuki, S. Fukami, A. Fujimori, H. Ohno, and M. Hayashi, "Giant perpendicular magnetic anisotropy in Ir/Co/Pt multilayers" *Phys. Rev. Mater.* **3**, 104419 (2019).

[7] Y. Ishikuro, M. Kawaguchi, N. Kato, Y.-C. Lau, and M. Hayashi, "Dzyaloshinskii-Moriya interaction and spin-orbit torque at the Ir/Co interface" *Phys. Rev. B* **99**, 134421 (2019).

[8] T. Fache, H. S. Tarazona, J. Liu, G. L'Vova, M. J. Applegate, J. C. Rojas-Sanchez, S. Petit-Watelot, C. V. Landauro, J. Quispe-Marcatoma, R. Morgunov, C. H. W. Barnes, and S. Mangin, "Nonmonotonic aftereffect measurements in perpendicular synthetic ferrimagnets" *Phys. Rev. B* **98**, 064410 (2018).

[9] B. Böhm, L. Fallarino, D. Pohl, B. Rellinghaus, K. Nielsch, N. S. Kiselev, and O. Hellwig, "Antiferromagnetic domain wall control via surface spin flop in fully tunable synthetic antiferromagnets with perpendicular magnetic anisotropy" *Phys. Rev. B* **100**, 140411(R) (2019).

[10] A. Hoffmann, "Spin Hall Effects in Metals" *IEEE Trans. Magn.* **49**, 5172 (2013).

[11] L. Liu, C.-F. Pai, Y. Li, H. W. Tseng, D. C. Ralph, and R. A. Buhrman, "Spin-Torque Switching with the Giant Spin Hall Effect of Tantalum" *Science* **336**, 555 (2012).

[12] A. Brataas, A. D. Kent, and H. Ohno, "Current-induced torques in magnetic materials" *Nat. Mater.* **11**, 372 (2012).

[13] T. Jungwirth, J. Wunderlich, and K. Olejník, "Spin Hall effect devices" *Nat. Mater.* **11**, 382 (2012).

[14] I. Mihai Miron, K. Garello, G. Gaudin, P.-J. Zermatten, M. V. Costache, S. Auffret, S. Bandiera, B. Rodmacq, A. Schuhl, and P. Gambardella, "Perpendicular switching of a single ferromagnetic layer induced by in-plane current injection" *Nature* **476**, 189 (2011).

[15] T. Kimura, Y. Otani, T. Sato, S. Takahashi, and S. Maekawa, "Room-Temperature Reversible Spin Hall Effect" *Phys. Rev. Lett.* **98**, 156601 (2007).

[16] J. E. Hirsch, "Spin Hall Effect" *Phys. Rev. Lett.* **83**, 1834 (1999).

[17] W. Zhang, M. B. Jungfleisch, W. Jiang, J. Sklenar, F. Y. Fradin, J. E. Pearson, J. B. Ketterson, and A. Hoffmann, "Spin pumping and inverse spin Hall effects—Insights for future spin-

[18] T. White, T. Bailey, M. Pierce, and C. W. Miller, "Strong Spin Pumping in Permalloy-Iridium Heterostructures" *IEEE Magn. Lett.* **8**, 1 (2017).

[19] Y. Tserkovnyak, A. Brataas, G. E. W. Bauer, and B. I. Halperin, "Nonlocal magnetization dynamics in ferromagnetic heterostructures" *Rev. Mod. Phys.* **77**, 1375 (2005).

[20] H. J. Jiao and G. E. W. Bauer, "Spin Backflow and ac Voltage Generation by Spin Pumping and the Inverse Spin Hall Effect" *Phys. Rev. Lett.* **110**, 217602 (2013).

[21] L. Bai, P. Hyde, Y. S. Gui, C.-M. Hu, V. Vlaminck, J. E. Pearson, S. D. Bader, and A. Hoffmann, "Universal Method for Separating Spin Pumping from Spin Rectification Voltage of Ferromagnetic Resonance" *Phys. Rev. Lett.* **111**, 217602 (2013).

[22] E. Saitoh, M. Ueda, H. Miyajima, and G. Tatara, "Conversion of spin current into charge current at room temperature: Inverse spin-Hall effect" *Appl. Phys. Lett.* **88**, 182509 (2006).

[23] O. Mosendz, V. Vlaminck, J. E. Pearson, F. Y. Fradin, G. E. W. Bauer, S. D. Bader, and A. Hoffmann, "Detection and quantification of inverse spin Hall effect from spin pumping in permalloy/normal metal bilayers" *Phys. Rev. B* **82**, 214403 (2010).

[24] C. Hahn, G. de Loubens, O. Klein, M. Viret, V. V. Naletov, and J. Ben Youssef, "Comparative measurements of inverse spin Hall effects and magnetoresistance in YIG/Pt and YIG/Ta" *Phys. Rev. B* **87**, 174417 (2013).

[25] K. Kondou, H. Sukegawa, S. Mitani, K. Tsukagoshi, and S. Kasai, "Evaluation of Spin Hall Angle and Spin Diffusion Length by Using Spin Current-Induced Ferromagnetic Resonance" *Appl. Phys. Express* **5**, 073002 (2012).

[26] J.-C. Rojas-Sánchez, N. Reyren, P. Laczkowski, W. Savero, J.-P. Attané, C. Deranlot, M. Jamet, J.-M. George, L. Vila, and H. Jaffrès, "Spin Pumping and Inverse Spin Hall Effect in Platinum: The Essential Role of Spin-Memory Loss at Metallic Interfaces" *Phys. Rev. Lett.* **112**, 106602 (2014).

[27] K. Ando, S. Takahashi, J. Ieda, Y. Kajiwara, H. Nakayama, T. Yoshino, K. Harii, Y. Fujikawa, M. Matsuo, S. Maekawa, and E. Saitoh, "Inverse spin-Hall effect induced by spin pumping in metallic system" *J. Appl. Phys.* **109**, 103913 (2011).

[28] L. Liu, T. Moriyama, D. C. Ralph, and R. A. Buhrman, "Spin-torque ferromagnetic resonance induced by the spin Hall effect" *Phys. Rev. Lett.* **106**, 036601 (2011).

[29] A. Azevedo, L. H. Vilela-Leão, R. L. Rodríguez-Suárez, A. F. Lacerda Santos, and S. M. Rezende, "Spin pumping and anisotropic magnetoresistance voltages in magnetic bilayers: Theory and experiment" *Phys. Rev. B* **83**, 144402 (2011).

[30] Z. Feng, J. Hu, L. Sun, B. You, D. Wu, J. Du, W. Zhang, A. Hu, Y. Yang, D. M. Tang, B. S. Zhang, and H. F. Ding, "Spin Hall angle quantification from spin pumping and microwave photoresistance" *Phys. Rev. B* **85**, 214423 (2012).

[31] H. Nakayama, K. Ando, K. Harii, T. Yoshino, R. Takahashi, Y. Kajiwara, K. Uchida, Y. Fujikawa, and E. Saitoh, "Geometry dependence on inverse spin Hall effect induced by spin pumping in Ni81Fe19/Pt films" *Phys. Rev. B* **85**, 144408 (2012).

[32] C. T. Boone, H. T. Nembach, J. M. Shaw, and T. J. Silva, "Spin transport parameters in metallic multilayers determined by ferromagnetic resonance measurements of spin-pumping" *J. Appl. Phys.* **113**, 153906 (2013).

[33] M. Obstbaum, M. Härtinger, H. G. Bauer, T. Meier, F. Swientek, C. H. Back, and G. Woltersdorf, "Inverse spin Hall effect in Ni81Fe19 normal-metal bilayers" *Phys. Rev. B* **89**, 060407(R) (2014).

[34] V. Castel, N. Vlietstra, J. Ben Youssef, and B. J. van Wees, "Platinum thickness dependence of the inverse spin-Hall voltage from spin pumping in a hybrid yttrium iron garnet/platinum system" *Appl. Phys. Lett.* **101**, 132414 (2012).

[35] O. D'Allivy Kelly, A. Anane, R. Bernard, J. Ben Youssef, C. Hahn, A. H. Molpeceres, C. Carrétéro, E. Jacquet, C. Deranlot, P. Bortolotti, R. Lebourgeois, J. C. Mage, G. De Loubens, O. Klein, V. Cros, and A. Fert, "Inverse spin Hall effect in nanometer-thick yttrium iron garnet/Pt system" *Appl. Phys. Lett.* **103**, 082408 (2013).

[36] C. Kittel, "Ferromagnetic resonance" *J. Phys. Le Radium* **12**, 291 (1951).

[37] J.-M. Beaujour, D. Ravelosona, I. Tudosa, E. E. Fullerton, and A. D. Kent, "Ferromagnetic resonance linewidth in ultrathin films with perpendicular magnetic anisotropy" *Phys. Rev. B* **80**, 180415(R) (2009).

[38] C. Bilzer, T. Devolder, J.-V. Kim, G. Counil, C. Chappert, S. Cardoso, and P. P. Freitas, "Study of the dynamic magnetic properties of soft CoFeB films" *J. Appl. Phys.* **100**, 53903 (2006).

[39] X. Liu, W. Zhang, M. J. Carter, and G. Xiao, "Ferromagnetic resonance and damping properties of CoFeB thin films as free layers in MgO-based magnetic tunnel junctions" *J. Appl. Phys.* **110**, 33910 (2011).

[40] A. Conca, J. Greser, T. Sebastian, S. Klingler, B. Obry, B. Leven, and B. Hillebrands, "Low spin-wave damping in amorphous Co40Fe40B 20 thin films" *J. Appl. Phys.* **113**, 10 (2013).

[41] A. Conca, E. T. Papaioannou, S. Klingler, J. Greser, T. Sebastian, B. Leven, J. Lösch, and B. Hillebrands, "Annealing influence on the Gilbert damping parameter and the exchange constant of CoFeB thin films" *Appl. Phys. Lett.* **104**, 1 (2014).

[42] F. Xu, Q. Huang, Z. Liao, S. Li, and C. K. Ong, "Tuning of magnetization dynamics in sputtered CoFeB thin film by gas pressure" *J. Appl. Phys.* **111**, 07A304 (2012).

[43] C. Guillemard, S. Petit-Watelot, S. Andrieu, and

J.-C. Rojas-Sánchez, "Charge-spin current conversion in high quality epitaxial Fe/Pt systems : Isotropic spin Hall angle along different in-plane crystalline directions" *Appl. Phys. Lett.* **113**, 262404 (2018).

[44] A. Ruiz-Calaforra, T. Brächer, V. Lauer, P. Pirro, B. Heinz, M. Geilen, A. V. Chumak, A. Conca, B. Leven, and B. Hillebrands, "The role of the non-magnetic material in spin pumping and magnetization dynamics in NiFe and CoFeB multilayer systems" *J. Appl. Phys.* **117**, 163901 (2015).

[45] C. Swindells, A. T. Hindmarch, A. J. Gallant, and D. Atkinson, "Spin transport across the interface in ferromagnetic/nonmagnetic systems" *Phys. Rev. B* **99**, 064406 (2019).

[46] D. J. Kim, S. Il Kim, S. Y. Park, K. D. Lee, and B. G. Park, "Ferromagnetic resonance spin pumping in CoFeB with highly resistive non-magnetic electrodes" *Curr. Appl. Phys.* **14**, 1344 (2014).

[47] W. Skowroński, Ł. Karwacki, S. Ziętek, J. Kanak, S. Łazarski, K. Grochot, T. Stobiecki, P. Kuświk, F. Stobiecki, and J. Barnaś, "Determination of Spin Hall Angle in Heavy-Metal/Co-Fe-B-Based Heterostructures with Interfacial Spin-Orbit Fields" *Phys. Rev. Appl.* **11**, 024039 (2019).

[48] J. Sinova, S. O. Valenzuela, J. Wunderlich, C. H. Back, and T. Jungwirth, "Spin Hall effects" *Rev. Mod. Phys.* **87**, 1213 (2015).

[49] Y. Liu, Z. Yuan, R. J. H. Wesselink, A. A. Starikov, and P. J. Kelly, "Interface enhancement of gilbert damping from first principles" *Phys. Rev. Lett.* **113**, 207202 (2014).

[50] J. C. Rojas-Sánchez and A. Fert, "Compared Efficiencies of Conversions between Charge and Spin Current by Spin-Orbit Interactions in Two- and Three-Dimensional Systems" *Phys. Rev. Appl.* **11**, 054049 (2019).

[51] E. Grimaldi, V. Krizakova, G. Sala, F. Yasin, S. Couet, G. Sankar Kar, K. Garello, and P. Gambardella, "Single-shot dynamics of spin–orbit torque and spin transfer torque switching in three-terminal magnetic tunnel junctions" *Nat. Nanotechnol.* **15**, 111 (2020).